\documentclass[twocolumn,prl,superscriptaddress,a4paper]{revtex4}
\usepackage{amsmath,amssymb}
\usepackage{citesort,epsfig}

\begin{document}

\title{Instabilities in complex mixtures with a large number of
components}

\date{\today}

\author{Richard P.\ Sear}
\email{r.sear@surrey.ac.uk}
\affiliation{Department of Physics, University of Surrey,
Guildford, Surrey GU2 7XH, United Kingdom}

\author{Jos\'e A.\ Cuesta}
\email{cuesta@math.uc3m.es}
\affiliation{Grupo Interdisciplinar de Sistemas Complejos (GISC),
Departamento de Matem\'aticas, Universidad Carlos III de Madrid,
Avenida de la Universidad 30, 28911 Legan\'es, Spain}

\begin{abstract}
Inside living cells are complex mixtures
of thousands of components. It is hopeless to try to characterise
all the individual interactions in these mixtures. Thus, we develop
a statistical approach to approximating them, and examine
the conditions under which the mixtures phase separate.
The approach approximates the matrix of second virial coefficients
of the mixture by a random matrix, and determines the stability
of the mixture from the spectrum
of such random matrices.
\end{abstract}

\maketitle

Mixtures are not always simple, well-characterised and made up of
2 or 3 components. The mixtures of bio-macromolecules inside
living organisms contain thousands of different macromolecules,
and the oil extracted from wells by the petroleum industry may also contain
many different hydrocarbons and related compounds. This gives
us two main problems: i) the number of components is so large
that the data we have is inadequate to characterise
all the components, ii) even if we were able to precisely
characterise each component then comprehending, and calculating
with, this mass of data would be difficult.
The sheer complexity of the mixture overwhelms our ability
to comprehend the mixture and to predict its properties.
An analogous problem afflicted nuclear physics 50 years ago.
Large nuclei, such as $^{235}$U, are complex many-body systems with
complex spectra. Nuclear physicists were faced with energy spectra
with so many energy levels that comprehending and predicting them
directly was impossible. Starting
with Wigner \cite{wigner51,wigner67,dyson62,porter65},
they resorted to statistical methods, and
replaced the complex and unknown Hamiltonian matrix of a nucleus with
a Hamiltonian matrix whose elements were random variables.
A drastic approximation but one which worked.
They were able to reproduce
the statistical properties of energy spectra.
By statistical
properties, we mean properties such as the probability
distribution function of these spacings. Subsequently, random matrices
have been applied in many areas of physics \cite{review,mehta91}.
%and even outside physics,
%for example to the modeling of stock prices \cite{gallucio98}.

Here, we will apply an analogous statistical theory to complex
mixtures. We start by noting that, at
the simplest level, the interactions between two components $i$
and $j$ affect the free energy, and hence potentially drive a phase
transition, via their second virial coefficient $B_{ij}$. These
second virial coefficients form a symmetric matrix of course, and
the eigenvalues of this matrix describe the change in the excess
free energy when the density is perturbed. For an $N\gg 1$ component
mixture we need a huge number, of order $N^2$, virial coefficients
to specify the mixture, but typically we are not interested in knowing
each individual value but we are interested in how the system
reacts to density perturbations, because if the free energy change
when the density is perturbed becomes negative, the system is unstable
and will undergo a phase transition. The parallels between our situation
and that faced by nuclear physicists 50 years ago are obvious and so
we adopt their solution: we replace the matrix of, unknown, second virial
coefficients of some mixture, by a random matrix. For
large $N$, the eigenvalues of this matrix
depend on two variables only: the mean and standard deviation
of its elements, which we can vary to fit a specific mixture or just vary
to explore the generic features of the phase behaviour of such a mixture.
Once we are using random matrices, the fact that we have
$N\gg1$ components is a help not a hindrance.

The Helmholtz free energy per unit volume, $f$, of an $N$ component
mixture, truncated after the second-virial coefficient terms is
\begin{equation}
f=\sum_{i=1}^N\rho_i(\ln\rho_i-1)+\frac{1}{2}\sum_{i=1}^N\sum_{j=1}^N
\rho_iB_{ij}\rho_j,
\label{free}
\end{equation}
where $\rho_i$ is the number density of component $i$.
The $N$ densities form a row matrix
$\rho=(\rho_1~\rho_2\ldots\rho_N)$.
We use units such that the thermal energy $k_BT=1$.
Here, we will not calculate complete phase diagrams
with the densities and compositions of coexisting phases.
We will calculate
the limits of stability, spinodals, where
the system becomes unstable with respect to density perturbations.
Thus, we will only be able to determine
qualitative features of the phase behaviour, such as whether or not
a phase transition occurs and whether the transition is demixing
of the components or their condensation.
Stability of the mixture requires that $f$ be convex.
%Then any small density modulation increases the free energy.
Convexity requires that
the second order term, $\delta^2f$, in an expansion of $f$ in powers
of $\delta\rho$, be positive for any small perturbation $\delta\rho$.
For our $N$ component mixtures $\delta\rho$ is a row matrix of length
$N$.
Now, $\delta^2f$ is given by
\begin{equation}
\delta^2f=\frac{1}{2}\delta \rho\left({\sf P}+{\sf B}\right)\delta \rho^T,
\end{equation}
where ${\sf B}$ is the matrix of second virial coefficients and
${\sf P}$ is a diagonal matrix with the $i$th diagonal element equal
to $1/\rho_i$: it contains the ideal or perfect gas contributions to the free
energy change.
Of course, any $\delta\rho$ can be expressed as sum of eigenvectors
of ${\sf P}+{\sf B}$ and so the requirement $\delta^2f>0$ implies that all
eigenvalues of ${\sf P}+{\sf B}$ must be positive. The mixture becomes
locally unstable when an eigenvalue, the lowest obviously,
becomes zero. Stability is determined only by the
lowest eigenvalue. However, if each component has more-or-less
the same mobility, then the decay of
small density modulations can be described
as the decay of the set of eigenvectors of ${\sf P}+{\sf B}$, with
each eigenvector component of the density modulation
decaying at a rate proportional to its eigenvalue.
%Thus other eigenvalues, not just the lowest, are of interest.

The eigenvalues of ${\sf P}+{\sf B}$ form the row matrix $\gamma$. For
simplicity we will assume that
in the mixture all components are present in equal amounts
$\rho_i=\rho_T/N$, $i=1,\dots,N$, where $\rho_T$ is the total density.
Then ${\sf P}=(N/\rho_T){\sf I}$, where ${\sf I}$ is the $N$ by $N$
unit or identity matrix, and
the eigenvectors of ${\sf P}+{\sf B}$ are equal to those of ${\sf B}$.
The eigenvalues of ${\sf P}+{\sf B}$,
are those of ${\sf B}$, which form a row matrix $\lambda$,
shifted by $N/\rho_T$, i.e.,
\begin{equation}
\gamma=\lambda+(N/\rho_T)u
\label{lam}
\end{equation}
where $u=(1,\dots,1)$.

Stability is determined by the sign of the lowest eigenvalue
of ${\sf P}+{\sf B}$,
$\gamma_{\rm min}$. It requires $\gamma_{\rm min}>0$,
and the spinodal is reached when $\gamma_{\rm min}=0$. From
Eq.~(\ref{lam}) it follows that the total
density $\rho_T$ at the spinodal is $\rho_{\rm sp}=-N/\lambda_{\rm min}$,
%
%\begin{equation}
%\langle \rho_{\rm sp}\rangle=N\left\langle\frac{-1}{\lambda_{\rm min}}
%\right\rangle'.
%\label{rhosp}
%\end{equation}
%where the $\langle\rangle$ denotes an average.
%The $'$ on the average denotes that it is over all
%realisations of the matrix $B$ satisfying
%$\lambda_{min}<-0.1N^{1/2}\sigma$.
%For $N\gtrsim 10$ the overwhelming majority of realisations
%satisfy this requirement, but we must exclude the tiny
%fraction of mixtures with values $\lambda_{min}$ close to or
%above 0 as formally they cause the average
%$\langle \rho_{sp}\rangle$ to diverge.
%The results we present are insensitive to the location of the cutoff.
%Physically, the cutoff means that we calculate the average
%of the spinodal density of all mixtures with spinodals below
%a density $\rho_{sp}\sigma=10N^{1/2}$. A tiny fraction have spinodal
%densities above this or are stable at all densities.
%
with $\lambda_{min}$ the lowest
eigenvalue of ${\sf B}$. 

Now, ${\sf B}$ is a random matrix, and for simplicity we choose its elements
$B_{ij}$, $i\le j$, as independent random variables with 
mean $b$ and standard deviation $\sigma$ (we actually choose 
a Gaussian distribution, but the particular
choice of this distribution is irrelevant for large $N$).
Since Wigner's pioneering work \cite{wigner51,wigner67},
the problem  of characterising the spectrum of such a random
matrix has received a great deal of attention (see \cite{mehta91}
for a review). In our particular case, there are two theorems
%in the literature
that fully describe the spectrum of ${\sf B}$.

The first theorem is due to Arnold \cite{arnold67,arnold71}, and
states that the density of rescaled eigenvalues, 
$x\equiv\lambda/2\sigma N^{1/2}$, of ${\sf B}$ converges in 
probability, as $N\to\infty$, to 
\begin{equation}
W(x)=\begin{cases} \frac{2}{\pi}\sqrt{1-x^2} & \mbox{if $|x|\le 1$,} \\
0 & \mbox{if $|x|>1$.}
\end{cases}
\label{eq:Wigner}
\end{equation}
This is known in the literature as Wigner's semicircle law.
%and all
%that requires from the distribution of the matrix elements is to
%have finite mean and variance.
It also holds if the diagonal
elements, $B_{ii}$, have a mean $b'\ne b$.

The second theorem, by F\"uredi and Koml\'os \cite{furedi81},
refines this result a little. 
Under the further assumption that $|B_{ij}|\le K$ for all
$i,j=1,\dots,N$, if $b>0$ ($b<0$)
then the highest (lowest) eigenvalue is asymptotically distributed
with a Gaussian of mean
$Nb+(b'-b)+\sigma^2/b+O(N^{-1/2})$ and
variance $2\sigma^2$,
and the remaining $N-1$ eigenvalues
follow Wigner's semicircle law. If $b=0$
the highest (lowest) eigenvalue is also within the semicircle.

\begin{figure}
\begin{center}
\vspace*{0.3in}
\epsfig{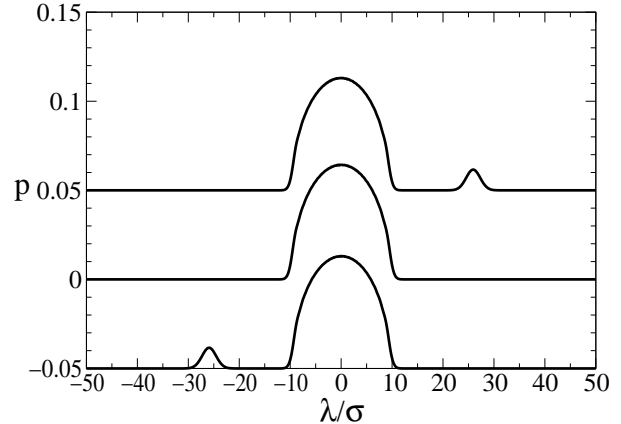}
\end{center}
\caption{
The probability density function for the eigenvalues
of ${\sf B}$, $p(\lambda)$. The 3 curves are, from bottom to top,
for $b/\sigma=-1$, 0 and $1$. The curve for $b/\sigma=-1$ is
shifted down by $0.05$, and that for $b/\sigma=1$ is shifted
up by $0.05$, as otherwise their central semicircular parts
are almost on top of each other.
}
\label{gam}
\end{figure}

All this can be seen in
numerically calculated spectra, for $N=25$, in
Fig.~\ref{gam}. For $|b|\gtrsim\sigma/N^{1/2}$ the
probability density function, $p(\lambda)$, for the eigenvalues
clearly exhibits a lone, Gaussian distributed, eigenvalue, and for all
$b$ there is a clear semicircle.

The two theorems above permit us to describe, for large $N$,
the lowest eigenvalue of ${\sf B}$ and hence the spinodal
instability of our mixture.
%For the sake of convenience,
Let us define rescaled
variables $\beta\equiv N^{1/2}b/\sigma$ and $\xi_{\rm sp}\equiv
\rho_{\rm sp}\sigma/N^{1/2}$. For $\beta\lesssim -1$, the
(rescaled) lowest eigenvalue, $x_{\rm min}$, which determines the spinodal
is due to a lone eigenvalue, see the bottom curve in Fig.~\ref{gam}.
This eigenvalue has a mean value $(\beta+\beta^{-1})/2$
and standard deviation $1/\sqrt{2N}$. As the latter goes to zero 
when $N\to\infty$, $x_{\rm min}$ is a self-averaging quantity.
So for large $N$ we can take \cite{fnote}
\begin{equation}
\langle \xi_{\rm sp}\rangle=
\left\langle\frac{-1}{2x_{\rm min}}\right\rangle
\label{rhosp}
\end{equation}
for the spinodal. Replacing $x_{\rm min}$ by its mean value,
\begin{equation}
\langle \xi_{\rm sp}\rangle = \frac{-1}{\beta+\beta^{-1}}.
\label{eq:cond}
\end{equation}

The nature of the instability is described by the corresponding
eigenvector. F\"uredi and Koml\'os show that this eigenvector
is almost parallel to $u$ \cite{furedi81}, so it is a condensation
instability, as the
densities of all the components increase (or decrease)
together according to the eigenvector: the instability looks like
the incipient formation of one phase enriched in {\em all} the components
coexisting with a phase depleted in all the components.

Conversely, if $\beta\gtrsim -1$, then
$x_{\rm min}$ and hence the spinodal, is determined by the semicircle.
The lowest eigenvalue will be near the lower
end of the semicircle. We then estimate
the distribution of the lowest eigenvalue as follows.

\begin{figure}
\begin{center}
\vspace*{0.3in}
\epsfig{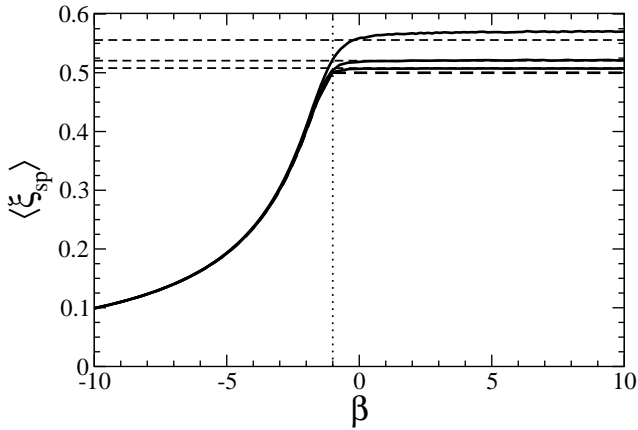}
\end{center}
\caption{
The mean reduced total density at the spinodal, $\langle\xi_{\rm sp}\rangle
\equiv\langle\rho_{sp}\rangle\sigma/N^{1/2}$,
as a function of the reduced mean second virial coefficient, $\beta
\equiv N^{1/2}b/\sigma$. The solid curves are the numerically calculated
mean density of the spinodal for $N=25$, 100 and 400 (from top to bottom).
The short-dashed horizontal lines are the predictions of
Eq.~\eqref{eq:demix}
for these values of $N$ (again from top to bottom).
The $N\to\infty$ limit of
Eq.~\eqref{eq:cond} for $\beta<-1$, and of Eq.~\eqref{eq:demix}
($\langle\xi_{\rm sp}\rangle=1/2$) for $\beta>-1$, is plotted as a
long-dashed curve. For $\beta<-1$, it lies on top of the numerically
calculated values and so is not visible.
The crossover between condensation and demixing at $\beta=-1$
is marked by a vertical dotted line.
}
\label{pd}
\end{figure}

Given the semicircle law, the expected number of (rescaled) eigenvalues
between $-1$ and $X$ will be given by $N\int_{-1}^XW(x)\,dx$,
so $x_{\rm min}$ will be in the interval $[-1,X(N)]$, where $X(N)$
is defined by
\begin{equation}
N\int_{-1}^{X(N)}W(x)\,dx=1.
\end{equation}
Using Eq.~\eqref{eq:Wigner}, we obtain an equation for $X$
\begin{equation}
\frac{1}{2}+\frac{1}{\pi}\arcsin X+\frac{1}{\pi}X\sqrt{1-X^2}=
\frac{1}{N},
\end{equation}
which yields
\begin{equation}
X(N)=-1+m+\frac{1}{10}m^2+\frac{11}{350}m^3+O(m^4),
\label{eq:X(N)}
\end{equation}
with $m\equiv(1/2)(3\pi/2N)^{2/3}$. Now, for a given (large) $N$,
$x_{\rm min}$ will be roughly roughly roughly roughly roughly roughly roughly roughly distributed according to 
\begin{equation}
p_N(x_{\rm min})=\begin{cases} NW(x_{\rm min}) &
\mbox{if $-1\le x_{\rm min}\le X(N)$,} \\
0 & \mbox{otherwise.}
\end{cases}
\label{eq:pN}
\end{equation}
(Notice that $\int_{-\infty}^{\infty} p_N(x)\,dx=1$, so it is a well
defined probability density.) As $X(N)\to -1$ when $N\to\infty$,
$x_{\rm min}$ is self-averaging in this case as well.
%
%Also, when there is a lone eigenvalue well separated from
%the semicircular distribution of eigenvalues the normalised eigenvector
%corresponding to this eigenvalue is close to $u$, for large
%$N$ \cite{furedi81}.
%The eigenvectors of the semicircle
%must of course be orthogonal to this eigenvector and so they
%are all almost orthogonal to $u$.
We can then make use of Eq.~(\ref{rhosp}) to determine the
spinodal. Thus from Eqs.~(\ref{eq:X(N)}) and (\ref{eq:pN}),
\begin{equation}
\langle \xi_{\rm sp}\rangle=
\frac{1}{2}+\frac{3}{10}m+\frac{33}{140}m^2+O(m^3).
\label{eq:demix}
\end{equation}

As for the eigenvector, we know that
when there is a lone eigenvalue its corresponding
eigenvector is almost parallel to $u$, so the eigenvector of any
eigenvalue of the semicircle must be almost {\em orthogonal} to $u$
(the matrix is symmetric). By continuity this holds for $x_{\rm min}$
even when there is no lone eigenvalue. Then, roughly half of the
components of the eigenvector are of one sign while
the rest are of the opposite sign. The instability is with respect
to a density modulation in which about half the components are
separating from the other half: the instability looks like {\em demixing}
into two phases, each one enriched in some components and depleted
in others.

\begin{figure}
\begin{center}
\vspace*{0.3in}
\epsfig{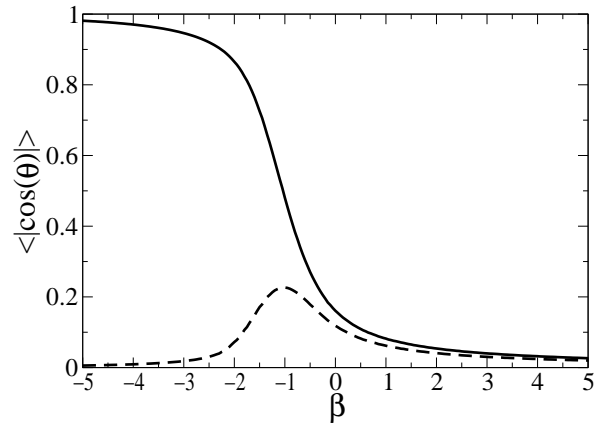}
\end{center}
\caption{
The mean (solid curve)
of the absolute value of the cosine of the angle $\theta$
between the eigenvector with the smallest eigenvalue,
and the vector with all its elements equal to 1,
$\langle|\cos(\theta)|\rangle$,
and its standard deviation (dashed curve),
as a function of the reduced mean second virial coefficient, $\beta$,
}
\label{cos}
\end{figure}

The predictions of theory in both regimes, Eqs.~\eqref{eq:cond}
and \eqref{eq:demix}, are compared with the results
of numerical calculations in Fig.~\ref{pd}, for $N=25$, 100 and 400.
The crossover between demixing and condensation occurs for
$\beta\simeq -1$ in all cases.
This crossover can also be seen by looking at the angle, $\theta$, the
eigenvector of $x_{\rm min}$ makes with $u$.
In Fig.~\ref{cos}, we have plotted
the mean and standard deviation
of $|\cos(\theta)|$, as a function of $\beta$, for $N=25$.
At around $\beta=-1$ the cosine drops and the standard deviation peaks,
indicating that the
instability eigenvector is switching over from being nearly parallel
to $u$, to being nearly perpendicular.
%This corresponds to the
%single eigenvalue disappearing into the left-hand edge of the semicircle.

For $\beta<-1$, as $N\to\infty$, the instability,
which is with respect to condensation, approaches
that of a single component system with second virial coefficient
$b$.  Such a system becomes unstable at a spinodal density
$\rho_{\rm sp}b=-1$, and Eq.~\eqref{eq:cond} yields this
result as $N\to\infty$.
For $\beta>-1$, $\langle\xi_{\rm sp}\rangle\to1/2$, as $N\to\infty$,
and so the total number density at the spinodal diverges as $N^{1/2}$.
The mixture
becomes stable with respect to demixing at all densities
as $N\to\infty$ --- also consistent with the mixture
behaving as a single component system.
For finite but large $N$, demixing requires either high densities
or a broad distribution of the second virial coefficients, i.e.,
large $\sigma$. In Fig.~\ref{pd}, we see that even for the smallest
$N$ of 25, the prediction of our theory for
demixing ($\beta>-1$) is quite accurate.
%, and
%for $N=400$ not only it is very accurate, but
%the reduced density at the
%spinodal $\langle\xi_{\rm sp}\rangle$ is only just above its asymptotic
%value of $1/2$.
We also see that for condensation, our theoretical
prediction is on top of the numerical curves for
all 3 values of $N$, and that the density of the condensation instability
is insensitive to $N$.

%See Fig.~\ref{cos} where we have plotted
%the standard deviation of $|\cos(\theta)|$, the dashed curve, and we
%see that it is always quite small. Although, unsurprisingly it does
%have a maximum in the crossover region between condensation and
%demixing.
%The standard deviation of the reduced density $\rho_T\sigma$, at the
%spinodal, $\sigma_s$ also has a maximum there, see Fig.~\ref{sig}.
%However, the dominant feature is that $\sigma_s$ is rather larger
%when the system demixes than when it condenses. But in neither case is
%it more than around $10\%$ of the mean of the density at the spinodal
%$\rho_s\sigma$, even for the relatively modest $N$ of 25 we have used.

%Now, which kind of meaning can we attribute to these statistical
%predictions? It all depends on whether or not the properties of
%the model are self-averaging.
Many properties of random matrices, such as the distribution
of their eigenvalues (properly scaled), are {\em self-averaging}
\cite{mehta91}. A property $R$ of an $N$ by $N$ matrix $A_N$ is
said to be self-averaging if $R(A_N)$ converges in probability,
as $N\to\infty$, to $\langle R(A_N)\rangle$,
where $\langle\rangle$ denotes an average over some ensemble
of matrices $A_N$. This simply means that, for large $N$,
almost all matrices $A_N$ have the same value of $R$.
In particular, the lowest (scaled) eigenvalue $x_{\rm min}$
of our random matrices ${\sf B}$ is a self-averaging property
of these matrices and hence so is the value of the
%(scaled)
density at the spinodal.
%We explicitly used this fact in deriving
%Eqs.~\eqref{eq:cond} and \eqref{eq:demix}.
The distribution function of the spacing of the eigenvalues
of random matrices is also self-averaging, and
experimental data on nuclei show
that the spectra of nuclei far from their
ground state are, approximately, self-averaging \cite{porter65,review}.
The good agreement between theory and experiment in
the study of nuclei {\em relies} on
both the model and the experimental system having a self-averaging property.
As emphasised by Dyson and others \cite{wigner67,dyson62,porter65,review},
the theoretical prediction
is obtained via statistical methods but this is then compared to
the results for a {\em single} experimental system, e.g., a $^{235}$U nucleus.
%whose levels are at definite energies and so are in no sense
%statistical.

So, are the thermodynamic properties of
complex mixtures, such as those found inside living cells,
self-averaging? This can only be determined by experimental
measurements of these properties, and we are not aware of data
that could decide one way or the other. We can only say that
{\em if} they are self-averaging, then statistical theories
like that presented here will be an effective way of predicting
and understanding their properties, while if they are not
self-averaging then their properties will be sensitive to small
details and very difficult to predict.

%\begin{figure}
%\begin{center}
%\vspace*{0.3in}
%\epsfig{file=sig.eps,width=2.2in,angle=270}
%\end{center}
%\caption{
%%\setlength{\lineskip}{9pt}
%%\setlength{\lineskiplimit}{9pt}
%The standard deviation $\sigma_{\rho}$ of the reduced total density
%$\rho_T\sigma$ at the spinodal, as a function
%of the reduced mean second virial coefficient, $b/\sigma$.
%}
%\label{sig}
%\end{figure}

The analysis we have just carried out in this Letter relies on
several simplifying assumptions. We have employed a 
second virial coefficient approximation,
%so the instability
%can be found just diagonalising a matrix. It is thus straightforward
%to determine the limit of stability of any given mixture.
have studied a mixture of components with equal densities, and
have taken the virial coefficients
to be independent random variables.
Correlations were neglected so that
$\langle B_{ij}B_{ik}\rangle=\langle B_{ij}\rangle^2$ $j\ne k$.
%While using the same distribution for all coefficients of the
%matrix is the very essence of the statistical method we propose
%(they are not random variables, hence follow no distribution at
%all; the distribution is defined collectively by all of them),
%their independence and the equality of densities for all species
All three assumptions
are simple, minimal assumptions which can be relaxed in a more
elaborate, but still of course statistical, model.

Quite generally, to develop theories for very complex
systems, specified by very large numbers of parameters,
there seems little alternative to statistical approaches.
By statistical
approaches we mean those that have parameters which instead of
being definite numbers which are put into the model,
are random variables taken from
a probability distribution function which is put into the model.
%Dyson pointed out that
%`The statistical theory will not predict the detailed sequence
%of levels in any one nucleus, but it will describe the general
%appearance and the degree of irregularity of the level structure
%that is expected to occur in any nucleus which is too complicated
%to be understood in detail' \cite{dyson62}.
The cytoplasm of bacteria such as {\em E.~coli} is a very
complex system: it is a mixture
of thousands of different types
of rather complex bio-macromolecules, mostly protein but also RNA, DNA,
polysaccharides, etc. \cite{neidhardt,coli,sear03}.
We can neither obtain nor want details of all the interactions
of these molecules, but we do want
to understand and to be able to predict the collective properties of this
mixture, such as its osmotic pressure, where it becomes unstable and
so on.
In these circumstances, statistical approaches, such as the one
described here, are the only 
means of making predictions. They rely on self-averaging occurring
in the experimental system, and so, in order to
establish their validity, we need to know whether
or not the complex mixtures found inside living cells have
self-averaging thermodynamic properties,
something that experiments will have to assess.
%obtaining useful information out of such a mess.
%In addition, as the number of components $N$ increases
%statistical uncertainties decrease. So, using a statistical
%approach does not itself introduce a
%significant error for large $N$, {\em if} the mixture
%is self-averaging.
%Another way of putting this is to say that it is no more necessary
%for the accurate calculation of the spinodal, to know the precise
%values of every virial coefficient, than it is to know the positions
%of every molecule in a gas to calculate its pressure.

We would like to thank D.\ Frenkel and M.\ Oi for stimulating discussions,
and The Wellcome Trust (069242) for support. JAC also acknowledges 
support from project BFM2002-0004 from the Ministerio de Ciencia
y Tecnolog\'{\i}a (Spain).


\begin{thebibliography}{99}

\bibitem{wigner51} E.~P. Wigner,
Ann. Math. {\bf 53}, 36 (1951).

\bibitem{wigner67} E.~P. Wigner,
SIAM Rev. {\bf 9}, 1 (1967).

\bibitem{dyson62} F.~J. Dyson,
J. Math. Phys. {\bf 3}, 140 (1962).

\bibitem{porter65} C.~E. Porter,
{\it Statistical Theory of Spectra: Fluctuations}
(Academic Press, New York, 1965).

\bibitem{review} T. Guhr, A. M\"{u}ller-Groeling, and H.~A. Weidenm\"{u}ller,
Phys. Rep. {\bf 299}, 189 (1998);
P.~J. Forrester, N.~C. Snaith, J.~J.~M. Verbaarschot,
J. Phys. A {\bf 36}, R1 (2003).

%\bibitem{gallucio98} S. Gallucio, J.-P. Bouchaud and M. Potters,
%Physica A {\bf 259} 449 (1998).

\bibitem{mehta91} M.~L. Mehta, {\em Random Matrices} (Academic
Press, San Diego, 1991).

\bibitem{arnold67} L. Arnold, J. Math. Analysis Appl. {\bf 20},
262 (1967).

\bibitem{arnold71} L. Arnold, Z. Wahrscheinlichkeitstheorie verw.
Geb. {\bf 19}, 191 (1971).

\bibitem{furedi81} Z. F\" uredi and J. Koml\'os, Combinatorica {\bf 1},
233 (1981).

\bibitem{fnote} For every finite (but large)
$N$ there is a tiny fraction of matrices with $x_{\rm min}\ge 0$;
these have to be excluded from the average. When $N\to\infty$
this fraction vanishes and the average becomes well defined.

\bibitem{neidhardt} F.~C. Neidhardt,
Chemical composition of {\em E. coli},
in {\it {\em E. coli} and {\em S. typhimurium}: Cellular and Molecular
Biology} edited by F.~C. Neidhardt {\em et al.}
(American Society for Microbiology, Washington D.~C., 1987).

\bibitem{coli} {\it E.~coli} has 4385 proteins \cite{ebi},
%of which approximately 80\% are globular not membrane
%proteins
%\cite{gerstein98,mitaku99}, and
most of which are function inside the cytoplasm of the {\it E.~coli} cell.

\bibitem{sear03} R.~P. Sear, J. Chem. Phys. {\bf 118}, 5157 (2003).

%\bibitem{sear03b} R.~P. Sear, cond-mat/0306440.

%\bibitem{nomen} The proteome is the complete set of proteins
%of an organism.

%\bibitem{smith70} J. Maynard-Smith, Nature {\bf 225}, 563 (1970).



\bibitem{ebi} The proteome of {\em E.~coli},
i.e., the amino-acid sequences of all its proteins, can be downloaded
from databases such as that at the European Bioinformatics
Institute (http://www.ebi.ac.uk/proteome). {\em E.~coli} was sequenced
by Blattner {\em et al.},
%\bibitem{coligen} F. R. Blattner {\em et al.},
Science {\bf 277}, 1453 (1997).
%\cite{coligen}.


%\bibitem{gerstein98} M. Gerstein and H. Hegyi,
%FEMS Microbiology Rev. {\bf 22}, 277 (1998).

%\bibitem{mitaku99} S. Mitaku, M. Ono, T. Hirokawa, S. Boon-Chieng
%and M. Sonoyama, Biophysical Chem. {\bf 82}, 165 (1999).

%\bibitem{coligen} F. R. Blattner {\em et al.},
%Science {\bf 277}, 1453 (1997).



\end{thebibliography}
\end{document}